\documentclass[conference]{IEEEtran}
\usepackage{amsmath,graphicx}
\usepackage{mathrsfs}
\usepackage{amssymb}
\usepackage{times,epsfig,float}
\usepackage{subfigure}
\usepackage{mathrsfs}
\usepackage{color}
\usepackage{times}
\usepackage{amsbsy}
\usepackage{latexsym}
\usepackage{amsmath}
\usepackage[ansinew]{inputenc}

\newcommand{\argmin}{\operatornamewithlimits{argmin}}
\newcommand{\argmax}{\operatornamewithlimits{argmax}}

\title{Adaptive Delay-Tolerant DSTBC in Opportunistic Relaying Cooperative MIMO Systems}

\author{Tong Peng and  Rodrigo C. de Lamare, \\
Communications Research Group, Department of Electronics, University
of York, York YO10 5DD, UK \\ CETUC/PUC-RIO, BRAZIL
\\
Email: tp525@ohm.york.ac.uk; rcdl500@ohm.york.ac.uk,}

\begin{document}
\maketitle
\begin{abstract}
An adaptive delay-tolerant distributed space-time coding (DSTC) scheme with feedback is proposed for two-hop cooperative multiple-input multiple-output (MIMO) networks using an amplify-and-forward strategy and opportunistic relaying algorithms. Maximum likelihood receivers and adjustable code matrices are considered subject to a power constraint. In the proposed delay-tolerant DSTC scheme, an adjustable code matrix is employed to transform the space-time coded matrices at the relay nodes. Stochastic gradient algorithms are developed with reduced computational complexity to estimate the parameters of the code matrix. Simulation results show that the proposed algorithms obtain significant performance gains and address the delay issue in cooperative MIMO systems as compared to existing DSTC schemes.
\end{abstract}
%
%
\section{Introduction}
\label{sec:intro}

Cooperative multiple-input multiple-output (MIMO) systems, which
employ multiple relay nodes with antennas between the source node
and the destination node as a distributed antenna array, can obtain
diversity gains by providing copies of the transmitted signals to
improve the reliability of wireless communication systems
\cite{Clarke}. Among the links between the relay nodes and the
destination node, cooperation strategies such as amplify-and-forward
(AF), decode-and-forward (DF), compress-and-forward (CF)
\cite{J.N.Laneman2004} and various distributed space-time coding
(DSTC) schemes \cite{J.N.Laneman2003}, \cite{Yiu S.}, \cite{RC De
Lamare} can be employed.

The development of distributed space-time coding (DSTC) schemes at
relay nodes in cooperative MIMO systems have been studied in order
to deal with asynchronous transmissions from the relays or delayed
reception of DSTC schemes at the destination has been addressed in
recent studies. By modifying the distributed threaded algebraic
space-time (TAST) codes \cite{Gamal}, Damen and Hammons designed a
delay-tolerant coding scheme in \cite{Damen} by the extension of the
Galois field employed in the coding scheme to achieve full diversity
and full rate. A further optimization which ensures that the codes
in \cite{Damen} obtain full diversity with the minimum length and
lower decoding complexity is presented in \cite{Torbatian}. The
authors of \cite{Zhimeng} proposed delay-tolerant distributed linear
convolutional STC schemes which can maintain the full diversity
property under any delay situation among the relays. In
\cite{Bhatnagar}, a transmit processing scheme based on linear
constellation precoder (LCP) design is employed to construct an
optimal DT-DSTC scheme to achieve the upper bound of the error
probability. In \cite{Behrouz Maham}, three opportunistic relaying
algorithms are designed for two-hop cooperative systems, named
opportunistic AF scheme, opportunistic source, and full-opportunism,
which can achieve a lower bit error rate (BER) performance compared
to the traditional equal power allocation schemes.

In this paper, we propose an adaptive delay-tolerant distributed
space-time coding scheme and algorithm for cooperative MIMO relaying
systems with feedback using opportunistic relaying algorithms with
different numbers of antennas at the relays. Two scenarios are
considered in the design: one in which the relays have multiple
antennas, and another in which the relays have a single antenna. We
propose a delay-tolerant adjustable code matrices opportunistic
relaying optimization (DT-ACMORO) algorithm based on the
maximum-likelihood (ML) criterion subject to constraints on the
transmitted power at the relays for different cooperative systems.
Specifically, stochastic gradient (SG) estimation methods
\cite{S.Haykin} are developed in order to compute the required
parameters at a reduced computational complexity. We study how the
adjustable code matrices affect the DSTC scheme during the encoding
process and how to optimize the adjustable code matrices by
employing an ML detector. The proposed delay-tolerant designs can be
implemented with different types of DSTC schemes in cooperative MIMO
relaying systems with DF or AF protocols.

The paper is organized as follows. Section II introduces two
different types of two-hop cooperative MIMO systems with multiple
relays applying the AF strategy and DSTC schemes. In Section III the
proposed optimization algorithms for the adjustable code matrices
are derived, and the results of the simulations are given in Section
IV. Section V gives the conclusions of the work.


\section{Cooperative MIMO System Model}

\begin{figure}
\begin{center}
\def\epsfsize#1#2{0.825\columnwidth}
\epsfbox{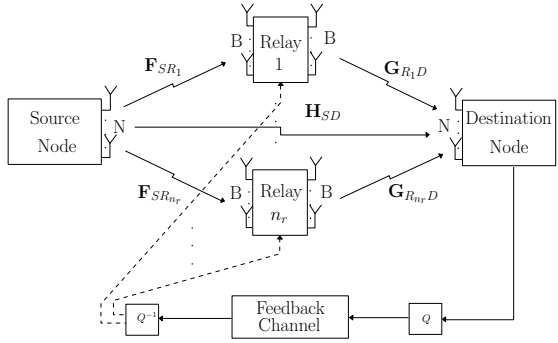}
\caption{Cooperative MIMO system model with
multiple-antenna $n_r$ relay nodes} \label{1}
\end{center}
\end{figure}

We consider a cooperative communication system, which consists of
one source node, $n_r$ relay nodes (Relay 1, Relay 2, ..., Relay
$n_r$), a feedback channel and one destination node as shown in Fig.
1. A two-hop cooperative MIMO system where all nodes employ $N\geq2$
antennas and each of them can either transmit or receive at one time
is considered. Two types of system configurations are considered in
this paper: one is called the multiple-antenna system (MAS)
configuration which employs relay nodes with $B=N$ antennas, and
another one is called the single-antenna system (SAS) configuration
which employs relay nodes with $B=1$ antenna. A delay profile
${\boldsymbol \Delta}=[\delta_1,\delta_2, ..., \delta_{n_r}]$ is
considered, where $\delta_k$ denotes the relative delay of the
signal received from the $k$th relay node as a reference to the
earliest received relay signal. Let ${\boldsymbol s}[i]$ denote the
transmitted information symbol vector at the source node, the
transmit power used at the source node is defined as $P_1$ and the
total power used among all the relay nodes is
$P_2=\sum_{k=1}^{n_rB}P_{2,k}$ where $k$ denotes the total number of
antennas among the relay nodes. We first consider the $k$th relay in
MAS. In the first phase, the source node will encode the MPSK or
M-QAM modulated information symbol vector ${\boldsymbol s}[i]$ to an
$T \times N$ STC matrix, which gives
\begin{small}
\begin{equation}
\label{2.1}
    {\boldsymbol S}[i] = \left[{\boldsymbol A}_1{\boldsymbol s}[i] ~{\boldsymbol A}_2{\boldsymbol s}[i] ~... ~{\boldsymbol A}_N{\boldsymbol s}[i]\right],
\end{equation}
\end{small}
where ${\boldsymbol A}_n$, $n=1,...,N$, are $T \times T$ unitary matrices. The source node broadcasts the STC matrix to the $n_r$ relay nodes after the encoding, and the received data at the $k$th relay node is given by
\begin{small}
\begin{equation}\label{2.2}
    {\boldsymbol X}_k[i] = \sqrt{\frac{P_1T}{N}}{\boldsymbol S}[i]{\boldsymbol F}_{MAS_k} + {\boldsymbol N}_k,
\end{equation}
\end{small}
where ${\boldsymbol F}_{MAS_k}$ denotes the $N \times N$ channel matrix between the source node and the $k$th relay node, and ${\boldsymbol N}_k$ is the matrix whose entries are complex zero-mean Gaussian random variables with variance $\sigma^2_{n_1}$.

In the second phase, the received symbols will be re-encoded and amplified before being forwarded to the destination node. It is worth to mention that instead of transmitting the fixed linear functions of the received matrix from the source node in \cite{Behrouz Maham}, we generate the adjustable code matrix ${\boldsymbol \Phi}_k[i]$ for the $k$th relay node and multiply it by the received symbol matrix before forwarding it to the destination node. The delay profile ${\boldsymbol \Delta}_k$ for each relay node is considered as well. The data communicated between the relays and the destination node can be described by
\begin{small}
\begin{equation}\label{2.3}
    {\boldsymbol R}[i] = \sum_{k=1}^{n_r}\sum_{j=1}^{N}\rho_{k,j}{\boldsymbol g}_{MAS_{k,j}}[i]{\boldsymbol y}_{k,j}[i]{\boldsymbol \Delta}_{k,j}[i] + {\boldsymbol N}_d,
\end{equation}
\end{small}
where $\rho_{k,j} = \sqrt{\frac{P_{k,j}}{\sigma^2_{{\boldsymbol F}_k}P_1+\sigma^2_{n_1}}}$, ${\boldsymbol y}_{k,j}[i] = {\boldsymbol \phi}_{k,j}[i]{\boldsymbol X}_k[i]$ and ${\boldsymbol \phi}_{k,j}[i]$ denotes the $j$th row in the $k$th adjustable code matrix, and ${\boldsymbol g}_{k,j}[i]$ is the $N \times 1$ channel coefficients vector between the $j$th antenna in the $k$th relay and the destination. The $N \times (\delta_{max}+T)$ delay profile matrix is denoted by ${\boldsymbol \Delta}_{k,j}$. By substituting (\ref{2.2}) into (\ref{2.3}) and rewriting the $(N \times (\delta_{max}+T))$ received matrix ${\boldsymbol R}[i]$ as an $(N(T+\delta_{max}) \times 1)$ vector we obtain
\begin{equation}\label{2.5}
\begin{aligned}
    {\boldsymbol r}[i] = & \sqrt{\frac{P_1T}{N}}\sum_{k=1}^{n_r}\sum_{j=1}^{N}\rho_{k,j}{\boldsymbol \Phi}_{MAS_{eq_{k,j}}}[i]{\boldsymbol H}^{\boldsymbol \Delta}_{MAS_{eq_{k,j}}}[i]{\boldsymbol s}[i]\\ & + \sum_{k=1}^{n_r}\sum_{j=1}^{N}\rho_{k,j}{\boldsymbol \Phi}_{MAS_{eq_{k,j}}}[i]{\boldsymbol G}^{\boldsymbol \Delta}_{MAS_{eq_{k,j}}}[i]{\boldsymbol n}_k + {\boldsymbol n}_d,
\end{aligned}
\end{equation}
where ${\boldsymbol H}^{\boldsymbol \Delta}_{MAS_{eq_{k,j}}}[i]$ is the $(N(\delta_{max}+T) \times N)$ equivalent channel matrix with the delay profile, and ${\boldsymbol \Phi}_{MAS_{eq_{k,j}}}[i]$ stands for the $(N(\delta_{max}+T) \times N(\delta_{max}+T))$ block diagonal adjustable code matrix. The noise vectors ${\boldsymbol n}_k$ and ${\boldsymbol n}_d$ are the columns of ${\boldsymbol N}_k$ and ${\boldsymbol N}_d$, respectively.

In the SAS configuration, the $N \times 1$ received symbols at the $k$th relay node are given by
\begin{equation}\label{2.7}
    {\boldsymbol x}_{SAS_k}[i] = \sqrt{\frac{P_1T}{N}}{\boldsymbol S}[i]{\boldsymbol f}_{SAS_k} + {\boldsymbol n}_{SAS_k},
\end{equation}
where ${\boldsymbol f}_{SAS_k}$ denotes the $N \times 1$ channel coefficients vector between the source node and the $k$th relay node, and the noise vector whose entries are complex zero-mean complex random variables with variance $\sigma_{n_1}$ is denoted by ${\boldsymbol n}_{SAS_k}$.

In the second phase, a $T \times T$ adjustable code matrix ${\boldsymbol \Phi}_{SAS_k}[i]$ is generated at the $k$th relay node and multiplied by the re-encoded symbol vector. The data transmitted between the relay nodes and the destination node is expressed as
\begin{equation}\label{2.8}
    {\boldsymbol R}_{SAS_k}[i] = \sum_{k=1}^{n_rN}\rho_k{\boldsymbol \Phi}_{SAS_k}[i]{\boldsymbol C}_k{\boldsymbol x}_{SAS_k}[i]{\boldsymbol g}_{SAS_k}{\boldsymbol \Delta}_k[i] + {\boldsymbol N}_d,
\end{equation}
where ${\boldsymbol C}_k$ stands for the $T \times T$ space-time code matrix at the $k$th relay node, and ${\boldsymbol g}_{SAS_k}$ denotes the $1 \times N$ channel coefficients vector between the $k$th relay and the destination. The matrix ${\boldsymbol N}_d$ has entries taken from complex Gaussian random variables with zero mean and variance $\sigma_d^2$. By substituting (\ref{2.7}) into (\ref{2.8}) we can obtain
\begin{equation}\label{2.10}
    {\boldsymbol R}[i] = \sqrt{\frac{P_1T}{N}}\sum_{k=1}^{n_rN}\rho_k{\boldsymbol \Phi}_{SAS_k}[i]{\boldsymbol C}_k{\boldsymbol S}[i]{\boldsymbol h}^{{\boldsymbol \Delta}}_k[i] + {\boldsymbol W},
\end{equation}
and we can rewrite the $(T \times (N+\delta_{max}))$ received matrix ${\boldsymbol R}[i]$ as an $T(N+\delta_{max}) \times 1$ vector given by
\begin{equation}\label{2.11}
\begin{aligned}
    {\boldsymbol r}[i] = & \sqrt{\frac{P_1T}{N}}\sum_{k=1}^{n_rN}\rho_k{\boldsymbol \Phi}_{SAS_{eq_k}}[i]{\boldsymbol H}^{\boldsymbol \Delta}_{SAS_k}[i]{\boldsymbol s}[i] \\ & + \sum_{k=1}^{n_rN}{\boldsymbol \Phi}_{SAS_{eq_k}}[i]{\boldsymbol G}^{\boldsymbol \Delta}_{SAS_k}[i]{\boldsymbol n}_{SAS_k} + {\boldsymbol n}_d,
\end{aligned}
\end{equation}
where ${\boldsymbol H}^{\boldsymbol \Delta}_{SAS_k}[i]$ and ${\boldsymbol G}^{\boldsymbol \Delta}_{SAS_k}[i]$ stand for the $T(\delta_{max}+N) \times T$ channel matrix with the delay profile incorporated into the symbol vector and the noise vectors, respectively. ${\boldsymbol \Phi}_{SAS_{eq_k}}[i]$ stands for the $T(\delta_{max}+N) \times T(\delta_{max}+N)$ diagonal equivalent adjustable code matrix. The use of an adjustable code matrix or a randomized matrix ${\boldsymbol \Phi}$ which achieves the full diversity order and provides a lower error probability has been discussed in \cite{Birsen Sirkeci-Mergen}. The uniform sphere randomized matrix which achieves the lowest BER of the analyzed schemes and contains elements that are uniformly distributed on the surface of a complex hyper-sphere of radius $\rho$ is used in our system. The delay profiles for different relay nodes are considered and the adjustable code matrices are normalized at the destination and then transmitted back to the relay nodes.

\section{Delay-Tolerant Adjustable Code Matrices Opportunistic Relaying Optimization}

In this section, we detail the proposed delay-tolerant adjustable code matrices opportunistic relaying optimization (DT-ACMORO) strategy with an ML receiver for designing DSTC schemes in MAS and SAS. Adaptive SG algorithms \cite{S.Haykin} for determining the parameters of the adjustable code matrix with reduced computational complexity are devised. A normalization is employed after the optimization so that no increase in the energy is introduced and the comparison between different schemes is fair. The optimized matrices will be sent back to the relays via the feedback channels. It is worth to mention that other adaptive estimation algorithms, such as RLS, can be used in the proposed scheme.

\subsection{DT-ACMORO Algorithm for MAS}

The constrained ML optimization problem for MAS can be written as
\begin{small}
\begin{equation}\label{3.0}
\begin{aligned}
     & \left[\hat {\boldsymbol s}[i], \hat {\boldsymbol \Phi}_{MAS_{eq_{k,j}}}[i]\right]  = \argmin_{{\boldsymbol s}[i],{\boldsymbol \Phi}_{MAS_{eq_{k,j}}}[i]} \|{\boldsymbol r}_{MAS}[i] - \bar {\boldsymbol r}_{MAS}[i]\|^2, \\
     & s.t.~ \sum_{k=1}^{n_r}\rm{Tr}({\boldsymbol \Phi}_{MAS_{eq_{k,j}}}[i]{\boldsymbol \Phi}^\emph{H}_{MAS_{eq_{k,j}}}[i])\leq \rm{P_R}, ~for~ d = 1,2,...,D,
\end{aligned}
\end{equation}
\end{small}
where $\bar {\boldsymbol r}_{MAS}[i]$ denotes the received symbol vector without noise which is determined by substituting each column of ${\boldsymbol S}$ into (\ref{2.5}). As mentioned in \cite{TARMO}, the optimization algorithm contains a discrete part which refers to the ML detection and a continuous part which refers to the optimization of the adjustable code matrices. The detection and the optimization can be implemented separately. As a result, other detectors such as MMSE and sphere decoders can be used in the detection part in order to reduce the computational complexity. The ML optimization problem for determining the transmitted symbol vector in (\ref{3.0}) is given by
\begin{small}
\begin{equation}\label{3.0.1}
\begin{aligned}
    &\hat {\boldsymbol s}[i] = \argmin_{{\boldsymbol s}_d\in{\boldsymbol S}}\\& \|{\boldsymbol r}_{MAS}[i]-\sqrt{\frac{P_1T}{N}}\sum_{k=1}^{n_r}\sum_{j=1}^{N}\rho_{k,j}{\boldsymbol \Phi}_{MAS_{eq_{k,j}}}[i]{\boldsymbol H}^{\boldsymbol \Delta}_{eq_{k,j}}[i]{\boldsymbol s}_d\|^2.
\end{aligned}
\end{equation}
\end{small}
By substituting each column in ${\boldsymbol S}$ into (\ref{3.0.1}) to search the most likely transmitted symbol vector $\hat {\boldsymbol s}[i]$ according to the ML detection rule, we can then calculate the optimal adjustable code matrix ${\boldsymbol \Phi}^{\boldsymbol \Delta}_{MAS_{eq_{k,j}}}[i]$.

The ML based Lagrangian expression is given by
\begin{equation}\label{3.1}
    \mathscr{L} = \|{\boldsymbol r}_{MAS}[i] - \sqrt{\frac{P_1T}{N}}\sum_{k=1}^{n_r}\sum_{j=1}^{N}\rho_{k,j}{\boldsymbol \Phi}_{eq_{k,j}}[i]{\boldsymbol H}^{\boldsymbol \Delta}_{eq_{k,j}}[i]\hat {\boldsymbol s}[i]\|^2.
\end{equation}
By taking the instantaneous gradient terms of the Lagrangian expression with respect to ${\boldsymbol \Phi}_{eq_{k,j}}[i]$ and equating the terms to zero, we can obtain
\begin{equation}\label{3.2}
     \nabla \mathscr{L}_{{\boldsymbol \Phi}_{eq_{k,j}}[i]} = -\hat {\boldsymbol s}^\emph{H}[i]({\boldsymbol H}^{\boldsymbol \Delta}_{eq_{k,j}}[i])^\emph{H}({\boldsymbol r}_{MAS}[i] - \hat{\boldsymbol r}[i]),
\end{equation}
where
\begin{equation}
    \hat{\boldsymbol r}[i]= \sqrt{\frac{P_1T}{N}}\sum_{k=1}^{n_r}\sum_{j=1}^{N}\rho_{k,j}{\boldsymbol \Phi}_{eq_{k,j}}[i]{\boldsymbol H}^{\boldsymbol \Delta}_{eq_{k,j}}[i]\hat {\boldsymbol s}[i].
\end{equation}
After we obtained (\ref{3.2}) the proposed SG algorithm can be obtained by introducing a step size into a gradient optimization algorithm to update the result until the convergence is reached. A detailed proof of the convergence of this type of algorithm is derived in \cite{UNiesen} and \cite{TARMO}. The SG algorithm is given by
\begin{equation}\label{3.3}
     {\boldsymbol \Phi}_{eq_{k,j}}[i+1] = {\boldsymbol \Phi}_{eq_{k,j}}[i] + \beta \hat {\boldsymbol s}^\emph{H}[i]({\boldsymbol H}^{\boldsymbol \Delta}_{eq_{k,j}}[i])^\emph{H}({\boldsymbol r}_{MAS}[i] - \hat{\boldsymbol r}[i]),
\end{equation}
where $\beta$ denotes the step size in the recursions for the estimation.

The opportunistic relaying problem in MAS can be expressed as
\begin{small}
\begin{equation}\label{3.4}
\begin{aligned}
    \hat k = & \argmax_{k=1,2,...,n_r}SNR_{k_{ins}}\sum_{j=1}^{N}\frac{P_1T\rho^2_{k,j}\|{\boldsymbol \Phi}_{eq_{k,j}}[i]{\boldsymbol H}^{\boldsymbol \Delta}_{eq_{k,j}}[i]\|^2_F\sigma^2_s}{N\rho^2_{k,j}\|{\boldsymbol \Phi}_{eq_{k,j}}[i]{\boldsymbol G}^{\boldsymbol \Delta}_{eq_{k,j}}[i]\|^2_F\sigma^2_{n_1}+\sigma^2_d},\\
    & ~~~~~~~~~~~~~~~~~~~\rm{for} ~ k=1,2,...,n_r.
\end{aligned}
\end{equation}
\end{small}
A summary of the DT-ACMORO SG algorithm in MAS is shown in Table I.

\begin{table}
  \centering
  \caption{Summary of the DT-ACMORO SG algorithm in MAS}\label{}
  \begin{tabular}{cc}
  \hline
1: & Initialize: generate ${\boldsymbol \Phi}[0]$ randomly with \\
   & the power constraint $\rm{Tr}({\boldsymbol \Phi}_{eq_k}{\boldsymbol \Phi}^\emph{H}_{eq_k})\leq \rm{P_R}$. \\
   \vspace{0.1cm}
2: & For each instant of time, $i$=1, 2, ..., compute \\
 & $\nabla \mathscr{L}_{{\boldsymbol \Phi}_{eq_{k,j}}[i]} = -\hat {\boldsymbol s}^\emph{H}[i]({\boldsymbol H}^{\boldsymbol \Delta}_{eq_{k,j}}[i])^\emph{H}({\boldsymbol r}_{MAS}[i] - \hat{\boldsymbol r}[i])$,\\
 \vspace{0.1cm}
3: & update ${\boldsymbol \Phi}_{eq_{k,j}}[i]$ using (\ref{3.3}),\\
\vspace{0.1cm}
4: & ${\boldsymbol \Phi}_{eq_{k,j}}[i] = \frac{\sqrt{\rm{P_R}}{\boldsymbol \Phi}_{eq_{k,j}}[i]}{\sqrt{\sum_{k=1}^{n_r}\rm{Tr}({\boldsymbol \Phi}_{eq_{k,j}}[i]({\boldsymbol \Phi}_{eq_{k,j}}[i])^\emph{H})}}$.\\
\vspace{0.1cm}
5: & Choose the optimal relay node by (\ref{3.4}) \\
   & and allocate all the transmit power to the optimal relay.\\
\hline
\end{tabular}
\end{table}

\subsection{DT-ACMORO Algorithm for SAS}

The main difference of the DT-ACMORO algorithm in SAS compared with that in MAS is the computational complexity which is related to the dimension of the adjustable code matrices used at the relays. We can rewrite the expression of the received vector for SAS in (\ref{2.11}) at the destination node as
\begin{equation}\label{3.10}
    {\boldsymbol r}_{SAS}[i] = \sum_{k=1}^{n_r}\sum_{j=1}^{N}{\boldsymbol D}^{\boldsymbol \Delta}_{eq_{k,j}}[i]\phi_{k,j}[i]{\boldsymbol s}[i] + {\boldsymbol n}_D[i],
\end{equation}
where ${\boldsymbol D}^{\boldsymbol \Delta}_{eq_{k,j}}[i]$ denotes the $T(\delta_{max} + N) \times N$ equivalent channel matrix combining the delay profile, the DSTC scheme, and the channel vector between the $j$th antenna of the $k$th relay node and the destination node. The quantity $\phi_{k,j}[i]$ denotes the adjustable code scalar assigned to the $j$th antenna the $k$th relay node. The ML based Lagrangian expression is written as
\begin{equation}\label{3.11}
    \mathscr{L} = \|{\boldsymbol r}_{SAS}[i] - \sum_{k=1}^{n_r}\sum_{j=1}^{N}{\boldsymbol D}^{\boldsymbol \Delta}_{eq_{k,j}}[i]\phi_{k,j}[i]\hat {\boldsymbol s}[i]\|^2,
\end{equation}
and a simple adaptive algorithm for determining the adjustable code matrices can be derived by taking the instantaneous gradient terms of the Lagrangian expression with respect to the scalar $\phi^*_{{k_j}_{SAS}}[i]$, which is
\begin{equation}\label{3.12}
     \nabla \mathscr{L}_{\phi^*_{k,j}[i]} = \nabla ({\boldsymbol r}_e[i])^\emph{H}_{\phi^*_{k,j}[i]}{\boldsymbol r}_e[i]
     = -{\boldsymbol s}^\emph{H}[i]({\boldsymbol G}^{\boldsymbol \Delta}_{eq_{k,j}}[i])^\emph{H}{\boldsymbol r}_e[i],
\end{equation}
where
\begin{equation}
    {\boldsymbol r}_e[i]={\boldsymbol r}_{SAS}[i] - \sum_{k=1}^{n_r}\sum_{j=1}^{N}{\boldsymbol G}^{\boldsymbol \Delta}_{eq_{k,j}}[i]\phi_{k,j}[i]\hat {\boldsymbol s}[i],
\end{equation}
stands for the error vector. After we arrived at (\ref{3.12}) the proposed SG algorithm can be obtained by introducing a step size into a gradient optimization algorithm to update the result until the convergence is reached, and the algorithm is given by
\begin{equation}\label{3.13}
     \phi_{k,j}[i+1] = \phi_{k,j}[i] + \beta ({\boldsymbol s}^\emph{H}[i]({\boldsymbol G}^{\boldsymbol \Delta}_{eq_{k,j}}[i])^\emph{H}{\boldsymbol r}_e[i]),
\end{equation}
where $\beta$ denotes the step size in the recursions for the estimation. The opportunistic relaying problem in SAS is given by
\begin{equation}\label{3.14}
\begin{aligned}
    \hat k = & \argmax_{k=1,2,...,n_r}\sum_{j=1}^{N}\frac{P_1T\rho^2_{k,j}\|{\boldsymbol D}^{\boldsymbol \Delta}_{eq_{k,j}}[i]\phi_{k,j}[i]\|^2_F\sigma^2_s}{N\rho^2_{k,j}\|{\boldsymbol G}^{\boldsymbol \Delta}_{eq_{k,j}}[i]\phi_{k,j}[i]\|^2_F\sigma^2_{n_1}+\sigma^2_d}, \\ &~~~~~~~~~~~~~~~~~~~\rm{for} ~ k=1,2,...,n_r.
\end{aligned}
\end{equation}
A summary of the DT-ACMORO SG algorithm in SAS is shown in Table II.

\begin{table}
  \centering
  \caption{Summary of the DT-ACMORO SG algorithm in SAS}\label{}
  \begin{tabular}{cc}
  \hline
1: & Initialize: generate ${\boldsymbol \Phi}[0]$ randomly with \\
   & the power constraint $\rm{Tr}({\boldsymbol \Phi}_{{eq_k}_{MAS}}{\boldsymbol \Phi}^\emph{H}_{{eq_k}_{MAS}})\leq \rm{P_R}$. \\
   \vspace{0.1cm}
2: & For each instant of time, $i$=1, 2, ..., compute \\
 & $\nabla \mathscr{L}_{\phi^*_{{k,j}_{SAS}}[i]} = -{\boldsymbol s}^\emph{H}[i]({\boldsymbol G}^{{\boldsymbol \Delta}_k}_{{eq_{k_j}}_{SAS}}[i])^\emph{H}{\boldsymbol r}_e[i]$,\\
 \vspace{0.1cm}
3: & update ${\boldsymbol \Phi}_{{eq_{k,j}}_{SAS}}[i]$ using (\ref{3.12}),\\
\vspace{0.1cm}
4: & ${\boldsymbol \Phi}_{{eq_{k,j}}_{SAS}}[i] = \frac{\sqrt{\rm{P_R}}{\boldsymbol \Phi}_{{eq_{k,j}}_{SAS}}[i]}{\sqrt{\sum_{k=1}^{n_r}\rm{Tr}({\boldsymbol \Phi}_{{eq_{k,j}}_{SAS}}[i]({\boldsymbol \Phi}_{{eq_{k,j}}_{SAS}}[i])^\emph{H})}}$.\\
\vspace{0.1cm}
5: & Choose the optimal relay node by (\ref{3.14}) \\
   & and allocate all the transmit power to the optimal relay.\\
\hline
\end{tabular}
\end{table}

\section{Simulations}

Simulation results are provided in this section to assess the proposed schemes and algorithms. The cooperative MIMO system considered employs the AF protocol with the distributed Alamouti (D-Alamouti) STBC scheme in \cite{RC De Lamare}, randomized Alamouti (R-Alamouti) in \cite{Birsen Sirkeci-Mergen} and linear dispersion code (LDC) in \cite{Hassibi_LDC} using BPSK and 4-QAM modulation in a quasi-static block fading channel with AWGN. The effect of the direct link is also considered. The cooperative MAS configuration equipped with $n_r=1,2$ relay nodes and $N=2$ antennas at each node, while the cooperative SAS scheme employed $N=2$ antennas at the source node and the destination node and $n_r=1,2$ relay nodes with a single antenna. In the simulations, we set the symbol power $\sigma^2_s$ as equal to 1, and the power of the adjustable code matrix in the DT-ACMO algorithm is normalized. The effect of the feedback is not considered in the simulations in this paper. Analysis and simulation results of the effects of limited feedback are considered in \cite{TARMO}.

The proposed DT-ACMORO algorithms with the Alamouti scheme and an ML receiver are evaluated with MAS in Fig. 2. Different delay profiles are considered in different relay nodes. It is shown in the figure that the BER result of the cooperative system with DSTC schemes and the AF protocol achieves a diversity order of $2$, and by employing opportunistic relaying (OR), fully-opportunistic (FO) and opportunistic source (OS) algorithms, a $2$dB to $3$dB BER improvement can be further achieved, respectively. In \cite{Behrouz Maham}, the benefits of employing three opportunistic relaying algorithms are explained. According to the simulation results in Fig. 2, a $2$dB to $3$dB gain can be achieved by using DT-ACMORO compared to the systems which only use opportunistic relaying algorithms. It is worth to mention that different DT-ACMORO algorithms require different computational complexities as shown in Table I, and the DT-ACMORO-FO algorithm which requires the choice of the best antenna at the source node and the use of the best relay node can achieve the best BER performance as compared with the DT-ACMORO-OR and DT-ACMORO-OS algorithms.
\begin{figure}
\begin{center}
\def\epsfsize#1#2{0.825\columnwidth}
\epsfbox{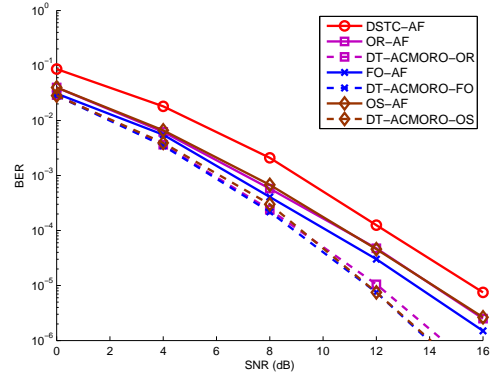}\vspace*{-1em} \caption{BER Performance vs. $SNR$
for DT-ACMORO Algorithm for MAS}\label{2} \vspace{-2em}
\end{center}
\end{figure}

In Fig. 3, the proposed DT-ACMORO SG algorithms are employed among the relay nodes in SAS. Compared to the opportunistic algorithms in \cite{Behrouz Maham}, the DT-ACMORO algorithms achieve a $2$dB to $4$dB improvement compared to the original opportunistic relaying algorithms as shown in Fig. 3. When comparing the curves in Fig. 3 to those in Fig. 2, it is noticed that the diversity order of the curves in Fig. 3 and that in Fig. 2 are the same because the same DSTBC schemes are employed in MAS and SAS, however the BER performances in Fig. 3 are worse that that in Fig. 2. The reason for worse BER performances in SAS is due to the lower number of antennas used in SAS according to the opportunistic relaying algorithms. In SAS $n_r=2$ relay nodes with a single antenna are employed and after the optimization, only one relay node with $N=1$ antenna is chosen to transmit the re-encoded information symbols. On the contrary, the best relay node in MAS contains $N=2$ antennas to forward the re-encoded DSTC scheme to the destination which ensures the complete DSTC scheme is received at the destination node.
\begin{figure}
\begin{center}
\def\epsfsize#1#2{0.825\columnwidth}
\epsfbox{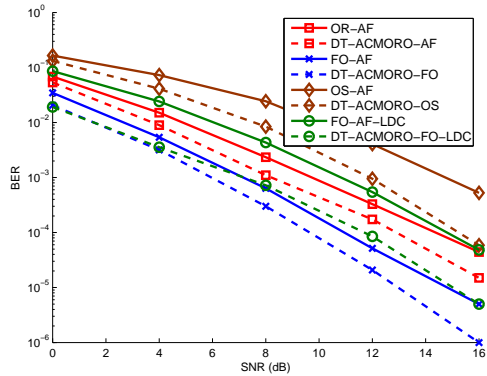}\vspace*{-1em} \caption{BER Performance vs. $SNR$
for DT-ACMORO Algorithm for SAS}\label{2} \vspace{-2em}
\end{center}
\end{figure}

\section{Conclusion}

We have proposed delay-tolerant adjustable code matrices
opportunistic relying optimization (DT-ACMORO) algorithms for
cooperative MIMO systems with feedback using an ML receiver at the
destination node to mitigate the effect of the delay associated with
DSTCs from relay nodes. Two types of cooperative systems, named MAS
and SAS, have been considered. Simulation results have illustrated
the performance of the two different cooperative systems with the
same number of transmit and receive antennas, and the advantage of
the proposed DT-ACMORO algorithms by comparing them with the
cooperative network employing the standard delay-tolerant DSTC
scheme. The proposed algorithms can be used with different DSTC
schemes and can also be extended to cooperative systems with any
number of antennas.

\bibliographystyle{IEEEbib}
\bibliography{strings,refs}

\bibliographystyle{IEEEtran}

\end{document}